\documentclass[conference]{IEEEtran}

\ifCLASSINFOpdf
  \usepackage[pdftex]{graphicx}

\else

\fi

\usepackage{amsmath,amssymb,amsthm}
\usepackage{algorithm,algorithmic}
\usepackage{bbm}
\usepackage{cite}
\usepackage{enumerate}
\usepackage{xspace}
\usepackage{todonotes}

\begin{document}
\bstctlcite{IEEEexample:BSTcontrol}
\title{Energy-Efficient Wireless Content Delivery with Proactive Caching}

\author{\IEEEauthorblockN{Samuel O. Somuyiwa, Andr\'as Gy\"orgy and Deniz G\"und\"uz}
\IEEEauthorblockA{Department of Electrical and Electronic Engineering, Imperial College London \\
Email: \{samuel.somuyiwa12, a.gyorgy, d.gunduz\}@imperial.ac.uk
}
}

\maketitle

\begin{abstract}
We propose an intelligent proactive content caching scheme to reduce the energy consumption in wireless downlink. We consider an online social network (OSN) setting where new contents are generated over time, and remain \textit{relevant} to the user for a random lifetime. Contents are downloaded to the user equipment (UE) through a time-varying wireless channel at an energy cost that depends on the channel state and the number of contents downloaded. The user accesses the OSN at random time instants, and consumes all the relevant contents. To reduce the energy consumption, we propose \textit{proactive caching} of contents under favorable channel conditions to a finite capacity cache memory. Assuming that the channel quality (or equivalently, the cost of downloading data) is memoryless over time slots, we show that the optimal caching policy, which may replace contents in the cache with shorter remaining lifetime with contents at the server that remain relevant longer, has certain threshold structure with respect to the channel quality. Since the optimal policy is computationally demanding in practice, we  introduce a simplified caching scheme and optimize its parameters using policy search. We also present two lower bounds on the energy consumption. We demonstrate through numerical simulations that the proposed caching scheme significantly reduces the energy consumption compared to traditional reactive caching tools, and achieves close-to-optimal performance for a wide variety of system parameters. 
\end{abstract}

\IEEEpeerreviewmaketitle

\newcommand{\cont}{K}
\newcommand{\live}{\mathcal{R}}
\newcommand{\cache}{\mathcal{I}}
\newcommand{\new}{\mathcal{N}}
\newcommand{\ind}[1]{\mathbb{I}_{\{#1\}}}
\newcommand{\req}{O}
\newcommand{\D}{D}  
\newcommand{\E}{\mathbb{E}}
\newcommand{\EE}[1]{\E\left[#1\right]}
\newcommand{\Pchar}{\mathbb{P}}
\newcommand{\PP}[1]{\Pchar\left[ #1 \right]}
\newcommand{\EEcond}[2]{\E\left[\left.#1\right|#2\right]}
\newcommand{\PPcond}[2]{\Pchar\left[\left.#1\right|#2\right]}
\newcommand{\X}{\mathcal{X}}
\renewcommand{\S}{\mathcal{S}}
\newcommand{\A}{\mathcal{A}}
\newcommand{\N}{\mathbb{N}}
\newcommand{\R}{\mathbb{R}}
\newcommand{\rel}{\mathcal{O}}
\newcommand{\New}{\mathcal{N}}
\newcommand{\Z}{\mathcal{Z}}
\newcommand{\G}{\mathcal{G}}
\newcommand{\paras}{{\theta}}
\newcommand{\para}{{\boldsymbol\paras}}
\newcommand{\thresh}{\mathcal{T}}

\newtheorem{theorem}{Theorem}
\newtheorem{lemma}{Lemma}
\newtheorem{definition}{Definition}
\newtheorem{proposition}{Proposition}
\newtheorem{corollary}{Corollary}

\vspace{-0.02cm}
\section{Introduction}
\label{sec:introduction}

Mobile data traffic is estimated to have a compound annual growth rate of 54\% between 2015--2020 \cite{Cisco:2017}. One of the major consequences of this growth is an increase in the energy consumption of wireless networks.
Emergence and growth of online social networks (OSNs), such as Facebook, Twitter, Instagram, play an important role in this escalating increase in mobile data traffic, as they enable a massive number of users to generate and share vast amounts of multimedia contents. 

Content caching has been proposed as a promising approach to address some of the challenges posed by the growing wireless data traffic. Among others, caching has been proposed to manage network traffic \cite{Proacseeding:2012}, reduce latency \cite{Chang:ICC:16}, reduce congestion \cite{FemtoCaching:INFOCOM}, and improve energy efficiency \cite{offlinecache:2015,Gregori:JSAC:16,zhou2016stochastic}. To improve the downlink energy efficiency of wireless networks, proactive caching of contents into a cache-enabled user equipment (UE) is proposed in \cite{offlinecache:2015,Gregori:JSAC:16}. These studies are done from an \textit{offline} optimization perspective by assuming that user demands and channel conditions are perfectly known in advance (non-causally). From a stochastic optimization perspective, \cite{zhou2016stochastic} studies how the energy efficiency of heterogeneous wireless network can be improved by exploiting the broadcast nature of wireless transmission to deliver cache contents to users. 

Most existing caching schemes do not take into account the dynamic variations in an OSN environment, and instead, cache contents are chosen based on a popularity profile, which typically is assumed to be static. However, in an OSN, contents remain popular only for a limited amount of time, and the popularity typically diminishes soon after the contents are generated
\cite{Socialmedia_1:2013}. For example, the average lifespan of a video posted on Facebook is about 2 hours, and it is even shorter on Twitter averaging 18 minutes \cite{M2Mconsulting:2016}. Due to this highly non-stationary behaviour of content popularity, it is shown in \cite{Socialmedia_2:2014} that there is a reduction in cache-hit ratio, i.e., the fraction of times a requested content is present in the local cache memory, and an increase in service delay when traditional caching schemes are used to cache social media contents. Moreover, most existing caching schemes in the literature assume a fixed content ``library.'' However, the dynamic nature of the OSN environment implies that the content library is time-varying as new contents are generated continuously, while some existing contents become irrelevant and forgotten. Therefore, there is a need to study proactive content caching taking into account the dynamic nature of content generation as well as the limited lifespan of the contents in an OSN environment. Together with the time-varying quality of the wireless channels and the limitations in cache capacities, it is necessary to design intelligent cache replacement policies that can adapt to this highly dynamic environment. 

In this paper, we consider proactive caching of social media contents into a cache-enabled user equipment (UE). We assume that the cache memory of the UE has a limited capacity, and our goal is to minimize the energy required to deliver contents to the UE over a time-varying wireless channel. We consider an OSN framework in which a random number of contents, in particular, high-rate video files, are posted continuously by a user's social connections. The most \textit{relevant} contents, which are determined by the OSN service provider, e.g., Facebook \cite{fb_newsroom:2016}, appear on the user's \textit{news feed}. We consider that each content has a random \textit{lifetime} that represents the time period it remains relevant to the user.  The wireless network operator delivers these contents to the user. At each time period of content delivery, the serving base station (BS) incurs a transmission energy cost that depends on the number of contents downloaded and the channel condition at that time period. We assume that the channel conditions are random and independent from one time period to another due to user mobility and/or variations in the environment and traffic conditions. We also assume that the user's behaviour in terms of accessing the OSN in every time period is also random. We consider that the user, upon accessing the OSN (e.g., refreshing the app on her mobile device), is interested in all the relevant contents at the time of access, and thus would like to download all the contents in her news feed. Whenever the user does not access the OSN, a cache manager (CM)\footnote{Depending on the protocol, CM can be located either at the BS or UE.} decides whether or not to pre-download and store contents in the cache memory of the UE, and whether or not to remove some of the contents from the cache memory (to create free space). Specifically, our problem can be viewed as a cache management problem under a completely dynamic and random environment.

Our objective is to \textit{minimize the long-term average energy cost of content delivery}, focusing exclusively on the energy cost of wireless transmission. We model this problem as an infinite-horizon Markov decision process (MDP) \cite{Bertsekas:book}.
We first prove that the optimal solution has a threshold structure with respect to the channel quality. This result does not follow from existing techniques in the literature as we allow continuous distributions for the channel quality. To overcome this technical difficulty, we define the concept of MDP with side information (MDP-SI). The optimal threshold values depend on the remaining lifetimes of the relevant contents as well as the elapsed time since the last user access. The classical dynamic programming (DP) tools are infeasible to be used in practice due to the exponentially large state space, and also because we consider undiscounted, average-case optimality \cite{Bertsekas:book}. As an alternative, we employ policy search techniques \cite{Policysearch}, which operate within a parameterised policy space. However, the parameter space for obtaining an optimal solution to this problem is still prohibitively large. Therefore, we propose a suboptimal proactive caching scheme, called \textit{Longest lifetime In--Shortest lifetime Out} (LISO), in which the threshold values depend only on the longest remaining lifetime of the contents outside the cache and the shortest remaining lifetime of the contents within the cache.

Finally, we compare the performance of LISO with two lower bounds. The first bound is obtained by assuming unlimited cache capacity (LB-UC), while the second is obtained by assuming non-causal knowledge of the user access times (LB-NCK). As benchmarks, we also consider a random caching scheme, and a reactive scheme that downloads contents only when the user accesses the OSN. Numerical simulations indicate the effectiveness of LISO in significantly reducing the energy consumption under a realistic wireless channel model.

\section{System Model}\label{s:system_model}

We consider a slotted time communication system. At the beginning of each time slot $t$, $M_t$ new contents, for example videos, of the same size\footnote{We can make this assumption without loss of generality assuming that videos are split into smaller frames of equal size.} are generated and added to the user's feed. We denote the set of newly generated contents by $\New_t$, that is, $M_t=|\New_t|$. The $i$-th content generated in time slot $t$  has a lifetime of $\cont_{t,i}$, that is, it remains relevant for the user for $\cont_{t,i}$ time slots. Equivalently, the user can request the $i$th content at any time slot starting at $t$, and no later than $t+\cont_{t,i}-1$. We assume that the user requests and consumes each content at most once (contents may expire even before the user would access them). We also assume that $\{M_t\}$ and $\{\cont_{t,i}\}$ are independent and identically distributed (i.i.d.) sequences.
We denote the set of relevant contents in the user's cache at the beginning of time slot $t$ by $\cache_t$, and those not in the cache (including the ones generated at the beginning of the time slot) by $\rel_t$. We denote by $U_t$ the decision of the user to access (or not) the contents in time slot $t$: If $U_t=1$, the user accesses the system and requests and consumes all the relevant contents at the time of access. That is, all the contents that are not already in the cache memory must be downloaded, and all the contents in the cache are moved to the application layer. 
If $U_t=0$, the user does not access the system, but the cache manager (CM) has the option to download and push some contents to the cache of the UE, and remove some others. 
Denoting the set of contents that are downloaded at time slot $t$ by $A^{(1)}_t$, and
the contents that are removed from the cache by $A^{(2)}_t$, we have $A^{(1)}_t= \rel_t$ and  $A^{(2)}_t= \cache_t$ if $U_t=1$, while $A^{(1)}_t \subset \rel_t$ and $A^{(2)}_t \subset \cache_t$, if $U_t=0$.~\footnote{Note that we explicitly exclude the strictly suboptimal behavior of downloading and removing a content within the same time slot without being consumed by the user.} 
Any content, whose lifetime expires at the end of a time slot is automatically removed from the cache. Finally, we assume that the user access sequence $\{U_t\}$ is an arrival process with i.i.d. inter-arrival times $\{D_n\}$, and let $D \ge 1$ denote a random variable with the generic distribution of inter-arrival times.

Due to our homogeneity assumption on the size of the contents, any content can be adequately described by its remaining lifetime. Therefore, the sets $\New_t$, $\rel_t$, $\cache_t$, $A^{(1)}_t$ and $A^{(2)}_t$ can be represented as multi-sets of remaining lifetimes (positive integers, with the set of all positive integer tuples denoted by $\N^*$). For simplicity, we will only talk about sets instead of multisets, or subsets instead of sub-multisets of multisets, and operations such as union should be treated in a multiset manner. For a multiset $Z$ with positive elements, let $Z-1=\{z > 0: z+1 \in Z\}$ denote the multiset obtained by reducing each element of $Z$ by $1$, and removing all elements which become $0$. Given this definition, if $U_t=0$ the dynamics of the sets $(\rel_t)$ and $(\cache_t)$ satisfy 
\begin{align*}
\rel_{t+1}&=\big((\rel_t \cup A^{(2)}_t \setminus A^{(1)}_t) -1\big) \cup \New_{t+1} \\
\cache_{t+1}&=(\cache_t \cup A^{(1)}_t \setminus A^{(2)}_t) -1,
\end{align*}
and if $U_t=1$, we have $\rel_{t+1}=\New_{t+1}$ and $\cache_{t+1}=\emptyset$. 

We assume that the UE is equipped with a cache memory of capacity $B$. This implies that the decision $A_t=(A^{(1)}_t,A^{(2)}_t)$ of the CM is constrained by the available cache capacity, and any valid decision leads to a new cache state with $|\cache_t|\le B$. 

Downloading a content at time $t$ has a cost $C_t$ that depends on the channel state at that time. Thus, at every time $t$, the BS incurs a cost $\mu_t = |A^{(1)}_t|\cdot C_t$, and the average energy cost for the BS after $T$ time slots is given by
\begin{equation}
\textstyle
J_T = \frac{1}{T}\sum_{t=1}^{T} |A^{(1)}_t| \cdot C_t.
\label{first_11}
\end{equation} 

We assume that $C_t \ge 0$ is a continuous random variable with cumulative distribution function (cdf) $F_C(c)$, i.i.d across time. We also assume that the sequences $\{C_t\}$,$\{M_t\}$,$\{K_{t,i}\}$,$\{D_n\}$ are independent of each other and are bounded by positive real numbers $C_{max},M_{max},K_{max},D_{max}$, respectively. Lastly, we assume that the CM is aware of the probability distributions governing the system, and its objective is to minimize the long term average energy cost given by
\begin{equation}
\label{utility_func}
\rho = \limsup_{T\to\infty} \EE{J_T}.
\end{equation}
\section{Optimal solution}

In this section we first describe how the above cache management problem can be modeled as a MDP, and then we show the structure of the optimal cache management policy.

\subsection{The MDP model}
A finite-state, finite-action MDP is characterized by a quadruple $(\S,\A,P,\mu)$, where $\S$ and $\A$, the state and action spaces, respectively, are finite sets, $P:\S \times \A \times \S \to [0,1]$ is a probability kernel, and $\mu:\S \times \A \to [0,\mu_{max}]\cup\{\infty\}$ is a cost function with some $\mu_{max}>0$. The purpose of introducing an infinite cost is to allow a different action set in every state without complicating notation: for every state $s \in \S$, the set $\A_s=\{a \in \A: \mu(s,a)<\infty\}$ denotes the set of possible actions (otherwise the agent suffers an infinite cost), and we assume that $\A_s \neq \emptyset$, $\forall s \in \S$. If the agent takes action $a \in \A$ in state $s \in \S$, we denote the transition probability to state $s' \in \S$ by $P(s'|s,a)$ (note that $\sum_{s' \in \S} P(s'|s,a)=1$ for all $s \in \S, a\in \A_s$). The goal is to minimize the long term average cost $\rho=\lim_{T\to\infty}\EE{\frac{1}{T}\sum_{t=1}^T \mu(S_t,A_t)}$. A deterministic policy $\pi:\S \to \A$ selects a single action for each state. Let $\Pi$ denote the set of all deterministic policies. 

For any policy $\pi \in \Pi$, let $P^\pi:\S\times\S \to [0,1]$ denote the transition kernel induced by $\pi$, that is, $P^\pi(s'|s)=P(s'|s,\pi(s))$. Assuming the Markov chain defined by $P^\pi$ is irreducible and aperiodic for all $\pi$,  and letting $\rho^\pi$ denote the average cost $\rho$ if $A_t=\pi(S_t)$, then there exists a deterministic policy $\pi^*$ that minimizes the infinite-horizon average cost \cite{Bertsekas:book}, such that 
\begin{equation}
\rho^{\pi^*} \triangleq \min_\pi\lim_{T\to\infty}\EE{\frac{1}{T}\sum_{t=1}^T \mu(S_t,\pi(S_t))},
\end{equation}
where the minimum is taken over all admissible (causal) control strategies.\footnote{Based on our assumption on $P^\pi$, the initial state does not matter, and the limit exists for any $\pi$.} 

For the cache management problem, at the end of time slot $t$, the state of the user can be described by the time elapsed since the last access, denoted by $E_t$. We assume user access at time $t=0$, i.e., $U_0=1$; then $E_t \triangleq \min\{t-n: 0\le n \le t, U_n=1\}$. We denote by $\X \subset \N^* \times \N^* \times \N$ the set of all possible combinations of $\rel_t, \cache_t$ and $E_t$, so that at time $t, X_t = (\rel_t, \cache_{t}, E_t) \in \X$ is the combined state of  the relevant contents out of the cache, relevant contents in the cache, and the elapsed time from the last user access. $\rel_t$ and $\cache_{t}$ are controllable variables that depend on $E_t$, which develops autonomously according to the distribution of $D$. The channel state $C_t \in \Z$ for some $\Z \subset \R$ is uncontrollable by the CM and develops autonomously according to its probability density function (pdf) $f_C(c), c\ge 0$. Given these, at time $t$, the state of the MDP can be described by
$S_t = (X_t,C_t) \in \S$, where $\S = \X \times \Z$. This leads to an MDP with uncountable state space, which is not straightforward to handle.

However, as described above, the state can be split into two parts: a controllable state $X_t \in \X$ and an uncontrollable state $C_t \in \Z$, which is i.i.d. and does not affect the set of available actions $A_{(X_t,C_t)}=A_{X_t}$; hence, $C_t$ essentially behaves as side information. This leads to the notion of a \textit{finite MDP with side information (MDP-SI)}, which we introduce in detail in the longer version of this paper \cite{Somuyiwa:Journal:2016}.

\subsection{Structure of the optimal policy}
In the MDP defined above, there is a set of available actions $\A_s$ for each pair $s=(x,c) \in \S$. Note that $\A_s$ does not depend on $c$, hence we can write $\A_x$ instead of $\A_s$, and the action can be represented as a function $\pi_x: \Z \to \A_x$, where we choose $\pi_x(c)$ in the original state $(x,c)$. In the MDP-SI setting we choose a function $\pi_x$ as an action in state $x \in \X$, leading to a finite state MDP where the transition probabilities and rewards are obtained as the expectation of the original transition probabilities and rewards, respectively, with respect to the channel distribution (pdf) $f_C$. In \cite{Somuyiwa:Journal:2016} we show that there is an optimal policy $\pi_x$ that is a piecewise constant function of the side information $c$. Therefore, the decision at time $t$ is a piecewise constant function of the channel cost $C_t$ with values taken from $\A_{X_t}$. To express $a \in \A_s$ more intuitively, we introduce the concept of a \textit{simple action}.

\begin{definition}
A simple action denoted by $a = (l|L)$, for $l \in \cache$,  and $L \in \rel$, is defined as follows: If $E_t = 0$, $a = (l|L)$ downloads the content represented by $L$ and removes the content represented by $l$ from the cache memory, and moves both contents to the application layer. If $E_t >0$, $a = (l|L)$ swaps the two contents. 
\end{definition}

Here, $l=0$ means that if $E>0$ (i.e., user does not access the contents), the content with lifetime $L$ is downloaded without removing any content from the cache memory, and if $E=0$ (i.e., user accesses the contents), the content is downloaded to the application layer without any corresponding removal of a content from the cache memory. Similarly, $L=0$ means that no actual download of content happens, while the content $l$ is removed from the cache. At every time slot $t$, because of the cache capacity constraint, the CM can only take up to $B$ simple actions if $E_t>0$. Therefore, any action of an optimal policy can be expressed as at most $B$ consecutive simple actions, and an action $a=(\{L_1,\ldots,L_{B'}\},\{l_1,\ldots,l_{B'}\})$,\footnote{If either $\vert \rel \vert$ or $\vert \cache \vert$ is less than $B'$, we simply zero-pad the set. } for some $0 \le B' \le B$, can be written as a sequence of simple actions $\{(l_1|L_1)\cdots(l_{B'}|L_{B'})\}$. We will next characterize the structure of the optimal policy in the following theorem (see \cite{Somuyiwa:Journal:2016} for a detailed proof).

\begin{theorem}
Consider the state $(\rel,\cache,E,C)$ for the MDP corresponding to this proactive cache management problem. Let $l_1\le \cdots \le l_B$ denote the contents in $\cache$, and let $L_1 \ge \cdots \ge L_B$ denote the $B$ largest elements in $\rel$. Then, there is a $B'\le B$, and corresponding threshold values $\thresh_{B'} < \thresh_{B'-1}<\cdots<\thresh_1<C_{max}$, such that, if $E>0$,\footnote{All policies behave exactly the same way when the user accesses a content; therefore, we only consider the case when this does not happen.} there is an optimal caching policy that performs the simple actions $(l_i|L_i)$ for all $1\le i\le B'$ for which $C\le \thresh_i$.
\label{theorem:thresh}
\end{theorem}

For an intuitive explanation of the theorem; given the differential value function for any state is defined as
\[
V^\pi(s) = \EEcond{\sum_{t=1}^\infty (\mu(S_t,A_t)-\rho^\pi)}{S_1=s},
\] 
the optimal policy $\pi^*$ in \eqref{utility_func} satisfies
\[
V^{\pi^*}(s) = \min_{a\in\A} \left\{\mu(s,a) + \sum_{s' \in \S} P^{\pi^*}(s'|s,a) V^{\pi^*}(s') \right\},
\]
such that $a=\pi^*(s)$ minimizes the right hand side. Notice that the value of a state increases if we replace a content $l \in \cache$ with another $L \in \rel$ with $L \ge l$. Therefore, the best action that performs exactly $B' \le B$ swaps (all such actions have the same cost) for $l_1\le \cdots \le l_B$ and $L_1 \ge \cdots \ge L_B$ is $A_{B'}\triangleq\{(l_1|L_1),\ldots,(l_{B'}|L_{B'})\}$. Therefore, an optimal policy $\pi^*$ can only use actions $A_1,\ldots,A_B$. If $b>b'$, for any $1< b \le B$, after applying $A_b$, the future value of the new state is at least as large as the value after applying $A_{b'}$; therefore, $A_{b'}$ can only be applied in an interval of the cost $C$ that is smaller than the interval for $A_b$. Using that $A_{b'}\subset A_b$, the structural result on the optimal policy follows.

Having proven that the optimal policy exhibits a threshold behavior, one still has to evaluate the optimal threshold values to characterize the optimal performance. However, the number of controlled states $(\rel,\cache,E)$ is enormous (roughly $O(B^{K_{max}} (M_{max} D_{max})^{K_{max}})$), making it infeasible to compute or store an optimal policy.
In the next section we describe a family of threshold policies with a simpler structure.

\section{LISO: A Suboptimal Policy}\label{s:policy_gradient}
The \emph{longest lifetime in--shortest lifetime out} (LISO) policy assigns a single threshold value to each pair of lifetimes, and swaps the content with the shortest remaining lifetime inside the cache with the content outside the cache that has the longest remaining lifetime, if the channel cost $C_t$ is below the threshold for this pair. This is repeated in each slot until no more swaps can be performed.\footnote{Note that contents with the shortest/longest lifetime in/out of the cache changes after each swap.}  
The LISO policy is parametrized by its threshold values $\para=(\thresh(a))_{a = (l|L)}$ for all pairs $a = (l|L)$ with $L > l$. Naturally,  $0 \le \thresh(a) \le C_{max}$ for all $a$. Furthermore, we can also impose monotonicity constraints $\thresh(l|L) \le \thresh(l|L')$ if $L'>L$, and $\thresh(l|L) \ge \thresh(l'|L)$ if $l'>l$.

While LISO significantly reduces the number of thresholds to be considered, especially when the cache capacity is large, computing the optimal thresholds is still challenging. As typical with learning parametrized policies over large state and action spaces (our policy space, representing the threshold vectors, is uncountable), we will resort to a policy search technique. In particular, we consider a Policy gradient method \cite{Policysearch}, which applies approximate gradient descent to minimize the expected average cost. where. Given a parameter vector $\para_j$ and corresponding policy $\pi_{\para_j}$, the parameter vector is updated along the gradient direction $\nabla_{\para} \rho^{\pi_{\para}}$ as follows
\vspace{-0.1cm}
\begin{align}
\para_{j+1} = \para_j - \lambda \nabla_{\para} \rho^{\pi_{\para}},
\label{polupdate}
\end{align}
where $\lambda>0$ is the step size. Let $P_{\para,T}$ denote the distribution of the trajectory of state-action pairs $\tau_{\pi_\para,T}=\big((S_1,A_1),\ldots,(S_T,A_T)\big)$ obtained by following the policy $\pi_{\para}$ for $T$ steps, and the corresponding average sample energy cost is given by $J_{\pi_{\para},T}(\tau_{\pi_{\para},T}) = \frac{1}{T}\sum_{t=1}^T \mu(S_t,A_t)$. For infinite trajectories ($T=\infty$) we shall use the simpler notation $P_\para, \tau_{\pi_\para}, J_{\pi_\para}$. Then the gradient of $\rho^{\pi_{\para}}$ can be expressed as
\vspace{-0.1cm}
\begin{align}
\nabla_{\para} \rho^{\pi_{\para}} = \int_\tau \nabla_{\para} P_{\para}(\tau_{\pi_\para})J_{\pi_\para}(\tau_{\pi_\para})d\tau~.
\label{polgrad}
\end{align}
Unfortunately, we do not have access to the true gradient vectors, however, under our assumptions, we have $\rho^{\pi}=J_{\pi}(\tau_\pi)$ with probability 1, and  $\rho^{\pi}=\lim_{T\to\infty}\mathbb{E}[J_{\pi,T}]$, thus, we can estimate the above quantities depending on infinite trajectories with sample averages over independent, finite trajectories (obtained via Monte Carlo rollouts). Next we describe a method for estimating the gradient $\nabla_{\para} \rho^{\pi_{\para}}$ 

\subsection{Finite Difference Method (FDM)}
In FDM, using the deterministic policy $\pi_{\para}$, we estimate the gradient from sample trajectories by applying small perturbations $\Delta{\para}^{[i]}$ (for trajectory $i$) to the current parameter vector ${\para}_j$, and evaluating the performance difference between the perturbed and unperturbed parameter vectors.  For each coordinate of $\para$, perturbations are drawn from a uniform distribution with range $(\Delta\theta_{min},\Delta\theta_{max}) = (-r,r)$, where $r\in\mathbb{R}^+$ is a relatively small number that must be selected carefully. For each perturbation (and trajectory), we obtain the performance difference estimate as $\Delta J^{[i]}_{\pi} = J_{\pi}(\para_j + \Delta{\para}^{[i]}) - J_{\pi}(\para_j)$. The gradient estimate can be obtained by regression as
\vspace{-0.1cm}
\begin{align}
\nabla_{\para} \rho^{\pi_{\para}} = \left(\Delta\boldsymbol\Theta^\top \Delta\boldsymbol\Theta\right)^{-1} \Delta\boldsymbol\Theta^\top \Delta\boldsymbol J_{\pi},
\end{align}
where $\Delta\boldsymbol\Theta = [\Delta{\para}^{[1]},\cdots, \Delta{\para}^{[N]}]^\top$ and $\Delta\boldsymbol J_{\pi}= [\Delta J^{[1]}_{\pi},\cdots, \Delta J^{[N]}_{\pi}]^\top$. When implementing the FDM, we impose the two monotonicity constraints mentioned earlier after every parameter perturbation and after every policy update. 

To curtail the effect of noise that is introduced due to the randomness of $J_{\pi,T}(\tau,T)$, we compute the gradient estimate multiple times from independent samples, and make a gradient step as in \eqref{polupdate} using the average of the estimated gradient.

\section{Performance Lower Bounds}\label{s:lower_bound}

We propose two lower bounds on the average cost. The first lower bound, called LB-UC, is obtained by considering an unlimited cache capacity. In this case there is no need to remove contents from the cache; therefore, the caching decision for each content that is not in the cache can be made individually, and independent of the contents of the cache. This gives rise to the following corollary.
\begin{corollary}
\label{corr_1}
Assume that the cache capacity is unlimited, that is, $B=\infty$. Then, for any state $x=(\rel,\cache,E)$ with $E>0$, there exist thresholds $0\le \thresh_1 \le \cdots \le \thresh_{K_{max}} \le C_{max}$ (recall that $K_{max}$ is the maximum lifetime) depending only on $E$, such that a content $L \in \rel$ is downloaded if $C \le \thresh_L$.
\label{Unlimited capacity}
\end{corollary}
The Corollary \ref{corr_1} follows from Theorem \ref{theorem:thresh} with $B=\infty$, and it shows that in this case, the optimal policy can be described by a single threshold for each remaining lifetime and $E$. Furthermore, if $E_t$ has a geometric distribution (i.e., the user accesses the OSN independently with the same probability in each time slot), the thresholds will depend only on the lifetime, but not on $E$ (due to its memoryless property). 

The second lower bound, called LB-NCK, is obtained by assuming non-causal knowledge of the user access times. In this case only the contents that will not expire by the time the user accesses the OSN are downloaded. Hence, we can simply consider a content queue of homogeneous items, to which (at most) $M_t$ new items are added at time $t$. At each time slot, we need to decide whether to download a content or not, depending on the channel state. If we decide to download, then, due to the homogeneity of the contents, we download as many contents as possible until the cache is full. Therefore, thresholds for LB-NCK can be evaluated for every \textit{time-to-user-access}, and the contents are downloaded if the instantaneous cost $C_t$ is less than the threshold, which becomes higher as we approach the user access time.

\vspace{-0.02cm}
\section{Numerical Results} \label{s:numerical}
In this section we present simulation results using the FDM algorithm to demonstrate the performance of the proposed LISO policy. We also compare its performance with a conventional reactive policy, which downloads contents at the time of access, a naive random caching policy, which caches each content with a constant probability until the cache is full, and the two lower bounds presented in Section \ref{s:lower_bound}.

We assume that the wireless channel between the UE and BS is subject to large scale fading due to path loss and shadowing, while the effect of small scale fading is averaged out through frequency/time/space diversity. We adopt the 3GPP channel model \cite{3GPP_release9}, and consider an urban micro (UMi) system, with a hexagonal cell layout and a non-line-of-sight (NLOS) scenario, such that the path loss is given by $PL(d) = 36.7\log_{10}(d) + 22.7 + 26\log{10}(f_c) + \mathcal{X}_\sigma$, where $d$ is the distance, in meters, between the serving BS and the user, $f_c$ is the center frequency in GHz, and $\mathcal{X}_\sigma$ is the zero-mean log-normal shadow fading parameter with standard deviation $\sigma = 4$ dB. In every time slot, $d$ is drawn from a uniform distribution between $50$ and $250$ meters, and $f_c = 2.5$, modeling a system in which the mobile user ends up in a different location in a different cell at each time slot. 

The instantaneous cost in time slot $t$, $C_t$, is obtained using Shannon's capacity formula, $R = W \log_2(1 + \mbox{SNR}),$ where $R$ is the transmission rate, $W$ is the channel bandwidth, and $\mbox{SNR} = P_{signal}/P_{noise}$ is the signal-to-noise ratio (assuming an interference free channel). We use system parameters consistent with the LTE network model \cite{Sesia:LTE:2011}; on a dB scale, the noise power is given by $P_{noise} = 10\log_{10}(kT) + 10\log_{10}W + NF$, where $kT = -174$ dBm/Hz is the noise power spectral density and $NF = 5$ dB is a typical noise figure. We obtain the cost as $C_t = P_{signal} - G_{TX} - G_{RX} + PL(d)$, where $G_{TX}$ and $G_{RX}$ are the transmit and receive antenna gains. We use the values $G_{TX} = 17$ dBi and $G_{RX} = 0$ dBi in our simulations. We assume a fixed bandwidth of $10$ MHz, and a spectral efficiency of $R/ W = 2$ bps/Hz. 

The random variable $M$ that generates $m_t$ new contents in every time slot is drawn from a uniform distribution over the set $\{1, 2, \ldots,M_{max}\}$. We also assume that the lifetimes $K_{t,i}$, ($1 \le i \le m_t$) of each new content generated in time slot $t$ is i.i.d. with uniform distribution over the set $\{5,10, \ldots, K_{max}\}$, where we set the maximum lifetime, $K_{max}$, as a multiple of $5$. We assume that the user accesses the network independently in each time slot with probability $p_a$. For all the simulations, we set the initial state as $\rel_0 = \cache_0 = \emptyset$.
 
For the FDM algorithm, the perturbation parameters are selected from $(\Delta\theta_{min}, \Delta\theta_{max}) = (-0.08, 0.08)$. For each policy update step, we use $\lambda=0.01$, and average five gradient estimates, each based on $100$ trajectories run for $300$ time slots. 
For performance tests, we use a test data of $100$ trajectories, each consisting of $5000$ time slots.

\begin{figure}
\centering
\includegraphics[width=0.9\columnwidth]{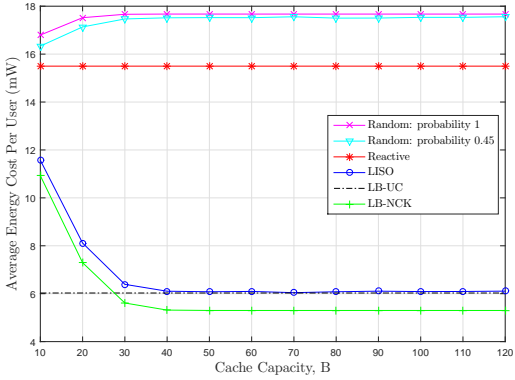}
\caption{Average energy cost vs. normalized cache capacity for different schemes when $K_{max} = 15, M_{max} = 8, D_{max} = 15, p_a = 0.25$.}
\label{C vs B}
\end{figure}

In Figure~\ref{C vs B} we illustrate the performance of LISO computed with the FDM algorithm for a finite capacity cache memory, together with the other schemes and lower bounds described above. Note that the instantaneous cost $C_t$ is in mW, and we simply assume that a time slot is normalized. We observe that the random caching scheme has the highest energy cost, which increases with the caching probability. Random caching scheme with zero caching probability is equivalent to reactive caching, and its performance degrades with increasing caching probability. This is because, both random caching and reactive schemes download contents at random time slots, resulting in the same energy cost on average; however, the random caching policy caches some files that are never requested by the user, whereas reactive caching downloads only requested contents. As expected, the average energy cost of reactive caching is independent of the cache capacity.

We see that the proposed LISO scheme significantly reduces the energy consumption with respect to these two schemes. More interestingly, in the low cache capacity regime, the performance of LISO closely follows the lower bound LB-NCK, obtained by assuming non-causal knowledge of the user access times; and for relatively higher cache capacities ($\sim B=30$), it approaches LB-UC. We have made similar observations with a wide range of system parameters, illustrating that the LISO scheme is close to optimal. Moreover, a relatively small cache capacity is sufficient to reap the benefits of proactive caching.



\begin{figure}
\centering
\includegraphics[width=0.9\columnwidth]{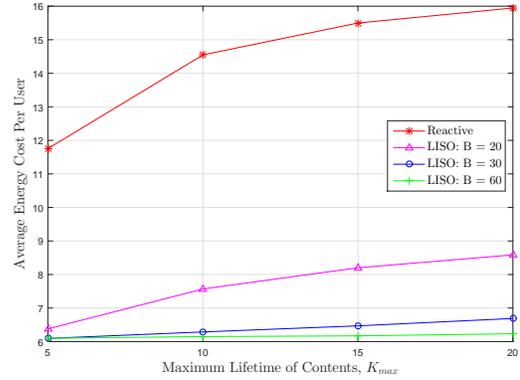}
\caption{Average energy cost vs. maximum lifetime of contents for the LISO scheme with cache capacities $B=20,30$ and $60$, and the reactive caching scheme, when $M_{max} = 8, D_{max} = 15, p_a = 0.25$.}
\label{C_vs_Kb}
\end{figure}

In Figure \ref{C_vs_Kb} we plot the average energy cost with respect to the maximum lifetime of contents. The result shows that the performance gain of LISO over reactive caching increases with the maximum lifetime. Average energy cost increases when contents remain relevant for a longer amount of time, because more files are requested and consumed by the user at the time of access, and therefore, more files should be cached. However, the increase in the energy cost (with increasing maximum lifetime) is much larger for reactive caching, as increased lifetime provides some additional flexibility to proactive caching to better exploit low-cost channel states. We also observe that the increase in the cache capacity becomes more useful as the maximum lifetime of the contents increases. 



\section{Conclusions}\label{s:conclusion}

We have studied proactive content caching to a mobile user to minimize the long term average energy cost. We considered an OSN setting with dynamic content generation, in which a random number of contents, each with a random lifetime, are generated at each time slot. The user accesses the OSN at random time intervals, and would like to consume all the relevant contents at the time of access. We first showed the optimality of a threshold-based caching policy, which replaces the contents in the cache depending on a threshold on the channel state, which depends on the remaining lifetimes of the relevant contents (in and out of the cache), and the time elapsed since last access. Since the space of policies with the optimal threshold structure is extremely large, we proposed a family of simpler policies with much fewer threshold values, which can be optimized much more efficiently. This family, called LISO, considers only the contents with the longest remaining lifetime outside the cache and with the shortest lifetime inside the cache to make a decision whether to swap the two contents or not. Finally, we exploited a policy gradient technique to compute the optimal threshold values. Our numerical simulations demonstrate that LISO can significantly outperform reactive caching, and can achieve close to optimal performance for a wide range of system settings.

\bibliographystyle{IEEEtran}
\bibliography{ccdwnref}
\end{document}